\documentclass[conference]{IEEEtran}
\usepackage{amssymb}
\usepackage{amsmath}
\usepackage[latin1]{inputenc}
\usepackage{mathabx}
\usepackage{algorithm}
\usepackage{algpseudocode}
\usepackage{pifont}
\usepackage{setspace}
\usepackage{psfrag}
\usepackage{color}
\usepackage{cite}
\usepackage{multirow}
\usepackage{wasysym}
\usepackage{graphicx,times,amsmath,balance,amsfonts,cite,pgf,tikz,pgffor}

\usepackage{pgfplots}
\usetikzlibrary{shadows.blur}
\usetikzlibrary{shapes.symbols}

\begin{document}
\title{Exploitation of Stragglers in Coded Computation}
\author{\IEEEauthorblockN{Shahrzad Kiani, Nuwan Ferdinand and Stark C. Draper}\\
	\IEEEauthorblockA{Department of Electrical and Computer  Engineering, University of Toronto, Toronto, ON, Canada}
	\IEEEauthorblockA{Email: shahrzad.kianidehkordi@mail.utoronto.ca, \{nuwan.ferdinand, stark.draper\}@utoronto.ca }}

\maketitle
\begin{abstract}
 In cloud computing systems slow processing nodes, often referred to as ``stragglers'', can significantly extend the computation time.  Recent results have shown
that error correction coding can be used to reduce the effect of stragglers. In this work we introduce a scheme that, in addition to using error correction to distribute mixed jobs across nodes, is also able to exploit the work completed by {\em all} nodes, including stragglers. We first consider vector-matrix multiplication and apply maximum distance separable (MDS) codes to small blocks of sub-matrices. The worker nodes process blocks sequentially, working block-by-block, transmitting partial per-block results to the master as they are completed. Sub-blocking allows a more continuous completion process, which thereby allows us to exploit the work of a much broader spectrum of processors and reduces computation time. We then apply this technique to matrix-matrix multiplication using product code. In this case, we show that the order of computing sub-tasks is a new degree of design freedom that can be exploited to reduce computation time further. We propose a novel approach to analyze the finishing time, which is different from typical order statistics. Simulation results show that the expected computation time decreases by a factor of at least two in compared to previous methods.
\end{abstract}

\section{Introduction}

The advent of large scale machine learning algorithms and data analytics has increased the demand for computation. Modern massive-scale computing tasks can no longer be solved using a single processor. Parallelization is required. There has been a recent surge in literature proposing different techniques to parallelize the fundamental computing primitives of machine learning and data analytics. Many approaches are tailored to specific algorithms with the general approach being a classic one, to decompose a computation task into a set of parallel sub-jobs. The number of sub-jobs determines the degree of acceleration. One such example is \emph{matrix multiplication}, a task found in many machine learning algorithms, e.g., sub-gradient calculations in stochastic gradient descent. As matrix multiplication can be decomposed into many small parallel jobs, it is possible to realize high degrees of parallelism.
 

In practical distributed computing environments the theoretical speedups promised will often not be attainable. Among other reasons, ``stragglers'' are a significant impediment to acceleration. Stragglers are \emph{slow workers}, who delay the computation of the final result. 
Recent work demonstrated that error correction coding (ECC) can be used to reduce the effect of straggler~\cite{Lee:ISIT16,Salman:unified,Lee:MATRIXISIT17,ferdinand:allerton16,Avestimehr:ISIT17, Cadambe:ISIT17,Alex2017:GC}. The central idea in~\cite{Lee:ISIT16} is to use maximum-distance separable (MDS) codes~\cite{Roth:2006} to generate redundant computations. 
The concept introduced in~\cite{Lee:ISIT16} has been extended in a number directions including matrix multiplication~\cite{Lee:MATRIXISIT17}, approximate computing~\cite{ferdinand:allerton16}, heterogeneous networks~\cite{Avestimehr:ISIT17} and convolution~\cite{Cadambe:ISIT17}.

One key feature of the coded computation approach in~\cite{Lee:ISIT16} (and all the papers that follow it) is that it ignores the work done by the worst $(n-k)$ nodes, nodes thereby deemed to be stragglers. In the case of {\em persistent} stragglers, i.e., worker nodes that are unavailable permanently or for an extremely long period, this is the ideal strategy. However, in practice, there are many \emph{non-persistent} stragglers, workers that, while slow, are able to do some amount of work. Non-persistent stragglers are present in practical cloud computing systems, and previous papers ignore the work they complete.  

In this paper, we propose a method to exploit the work completed by all workers, including stragglers. We first apply our coding scheme to vector-matrix multiplication. We decompose the matrices into much smaller sub-matrices, encode them using MDS codes, and assign each worker a set of subtasks. Each worker then sequentially computes subtasks. They \emph{transmit back} to the master the computed result of each subtask. I.e, a worker first computes its first subtask; transmits back the result before starting on the second subtask and so forth. The master node sequentially receives the completed subtasks from the workers. A faster worker may send a greater number of subtask results, while stragglers may send a smaller number. Once the master receives enough, it can recover the desired solution. We extend this method to matrix-matrix multiplication using product code. Through illustration we show that an ``order of processing" effect is pre-eminent in matrix-matrix multiplication, an effect that is not presented in the vector-matrix multiplication case. We then propose an order of processing  that reduces compute time.

In contrast to previous work, an important aspect of our model and results is that it leverages the sequential processing nature of most computing systems. In our paper, each worker sequentially processes multiple (small) encoded tasks in contrasts to processing a single (big) encoded task in~\cite{Lee:ISIT16,Lee:MATRIXISIT17}. This means that in our paper, the processing times of encoded tasks are no longer independent and identically distributed as they are in~\cite{Lee:ISIT16,Lee:MATRIXISIT17}. Thus, standard order statistics cannot be used to analyze the latency performance of our scheme as was done in~\cite{Lee:ISIT16,Lee:MATRIXISIT17}. To this end, we propose a novel theoretical approach to study the variation of work done across workers. Our analysis illustrate how our strategy improves finishing times through effective exploitation of the work completed by all workers. 
	

\section{Vector-Matrix Multiplication} \label{secvector}

In this section, we propose our straggler exploitation method for vector-matrix multiplication. We detail our
proposed scheme in three sub-sections: the delegation of work by the master, the computation
at the workers, and the combining operation at the
master. Finally, we give an example and compare our scheme to existing schemes.

\subsection{The delegation of work by the master}
\label{sec.A}
We consider a distributed computing environment that consists of a
master and $n$ workers. The objective of the master is to perform the
vector-matrix multiplication $\mathbf A \mathbf x$ where $\mathbf A$
is an $m\times q$ matrix and $\mathbf x$ is a $q\times 1$ vector. We
first partition $\mathbf A$ into $k$ equally-sized sub-matrices ($k$
is a parameter of our scheme): 
\[\mathbf A =
\begin{bmatrix}
\mathbf A_1 ;       \ \ 
\mathbf A_2 ;       \ \ 
\hdots \ \  ;
\mathbf A_k     
\end{bmatrix}.
\]
Each sub-matrix $\mathbf A_i$ is of size $m/k\times q$. We next define
an $L\times k$ matrix $\mathbf G$ in which any $k$ row vectors of
$\mathbf G$ are linearly independent and any square matrix formed
using any $k$ columns of $\mathbf G$ is invertible. These conditions
can be satisfied with high probability by selecting the elements of $\mathbf G$ in an
independent and identically distributed (i.i.d.) manner from the
Gaussian normal distribution.  Let $\mathbf I_{m/k}$ be the $m/k
\times m/k$ identity matrix. The master computes
\begin{align}
\label{eqn:encoding}
\mathbf { \Bar{A}}= \left(\mathbf G \otimes \mathbf I_{m/k}\right)\mathbf A 
\end{align}
where $\otimes$ denote the Kronecker product and $\mathbf { \Bar{A}}$ is an
$Lm/k \times q$ matrix. The matrix $\mathbf { \Bar{A}}$ is composed of $L$
distinct sub-matrices (each of size $m/k\times q$):
\[\mathbf { \Bar{A}}=
\begin{bmatrix}
\mathbf{ \Bar{A}}_1;       \ \ 
\mathbf { \Bar{A}}_2;       \ \ 
\hdots \ \ ;
\mathbf { \Bar{A}}_L     
\end{bmatrix},
\]
The matrix $\mathbf{ \Bar{A}_i}$ is a linear combination of the
$\mathbf A_j$:
\begin{align}
\mathbf{\Bar{A}}_i = \sum_{j=1}^{k} g_{ij} \mathbf A_j
\end{align}
where $g_{ij}$ is the $ij$-th element of $\mathbf G$. The master
transmits $l_i$ distinct sub-matrices to worker $i$ where
$\sum_{i=1}^nl_i=L$ and $l_i>1$. All sub-matrices are distributed to distinct workers, i.e., no single matrix is given to two workers. Finally, the master sends $\mathbf x$ to all
workers.
  
\subsection{The computation at workers}
\label{sec.B}
The $i$-th worker receives $\mathbf{ \Bar{A}}_{(i-1)L/n+1}, \ldots \mathbf
{\Bar{A}}_{iL/n}$. It first computes $\mathbf w _{(i-1)L/n+1}=\mathbf
{\Bar{A}}_{(i-1)L/n+1} \mathbf x$ and transmits the result $\mathbf w
_{(i-1)L/n+1}$ back to the master. That same worker next
computes $\mathbf w _{(i-1)L/n+2}=\mathbf{ \Bar{A}}_{(i-1)L/n+2}\mathbf x$
and sends the result to the master. Likewise, it sequentially computes
block-by-block up to $l_i$ blocks, transmitting each result to
the master. The transmission of partial (per-sub-block) results is a
novel aspect of our scheme and is an essential aspect required to
exploit the work performed by all workers.

\subsection{Combining operation at master}
\label{sec.C}

The master receives blocks sequentially from all workers. Once the
master has received any $k$ distinct sub-computations from any set of
workers, it combines them to find the final vector-matrix multiplication $\mathbf A \mathbf x$. Let $\mathcal I
\subset \{1,\ldots L\}$ be the indexes of the $k$ block received. To
recover the desired output the master computes
\begin{align}
\label{eqn:decoding}
\mathbf y = \left(\mathbf G (\mathcal I)^{-1}\otimes \mathbf I_{m/k}\right) \mathbf w.
\end{align}
The matrix $\mathbf G (\mathcal I)$ is a $k\times k$ sub-matrix of
$\mathbf G$ with rows selected based on $\mathcal I$, and $\mathbf w$
consists of the received computed sub-computations concatenated according to
the order of the indexes in $\mathcal I$.

\subsection{An example}
\label{sec:example}
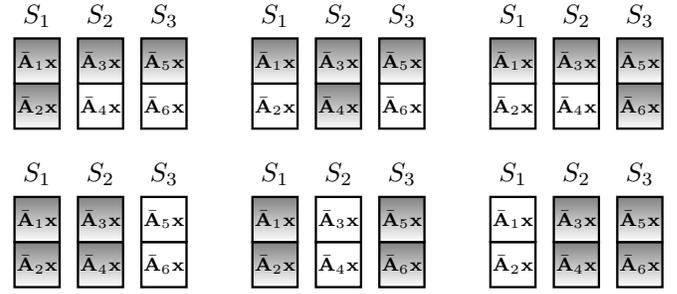
\begin{figure}
	\begin{tikzpicture}
	[scale=0.5,
	fine/.style = {solid,draw=gray},
	shaping/.style = {line width=0.3mm},
    shaping2/.style = {line width=0.25mm},
    shaping3/.style = {line width=0.4mm},
	voronoi/.style = {fill=blue!20!white}]
    \begin{scope}[scale=0.8]
    \begin{scope}
    \begin{scope}
		\begin{scope}[scale = 3,xshift=-45]
			\draw[shaping,shade] (0,0) rectangle (0.5,0.5);
			\draw[shaping,shade] (0,0.5) rectangle (0.5,1);
			\node at (0.25,0.25) {\scriptsize $\Bar{\mathbf A}_2 \mathbf x$};
			\node at (0.25,0.75) {\scriptsize $\Bar{\mathbf A}_1\mathbf x$};
			\node at (0.25,1.25) {$S_1$};
		\end{scope}
		
		\begin{scope}[scale = 3,xshift=-25]
			\draw[shaping] (0,0) rectangle (0.5,0.5);
			\draw[shaping,shade] (0,0.5) rectangle (0.5,1);
			\node at (0.25,0.25) {\scriptsize $\Bar{\mathbf A}_4\mathbf x$};
			\node at (0.25,0.75) {\scriptsize $\Bar{\mathbf A}_3\mathbf x$};
			\node at (0.25,1.25) {$S_2$};
		\end{scope}

		\begin{scope}[scale = 3,xshift=-5]
			\draw[shaping] (0,0) rectangle (0.5,0.5);
			\draw[shaping,shade] (0,0.5) rectangle (0.5,1);
			\node at (0.25,0.25) {\scriptsize $\Bar{\mathbf A}_6\mathbf x$};
			\node at (0.25,0.75) {\scriptsize $\Bar{\mathbf A}_5\mathbf x$};
			\node at (0.25,1.25) {$S_3$};
		\end{scope}		
	\end{scope}

    \begin{scope}[xshift=225]
		\begin{scope}[scale = 3,xshift=-45]
			\draw[shaping] (0,0) rectangle (0.5,0.5);
			\draw[shaping,shade] (0,0.5) rectangle (0.5,1);
			\node at (0.25,0.25) {\scriptsize $\Bar{\mathbf A}_2\mathbf x$};
			\node at (0.25,0.75) {\scriptsize $\Bar{\mathbf A}_1\mathbf x$};
			\node at (0.25,1.25) {$S_1$};
		\end{scope}
		
		\begin{scope}[scale = 3,xshift=-25]
			\draw[shaping,shade] (0,0) rectangle (0.5,0.5);
			\draw[shaping,shade] (0,0.5) rectangle (0.5,1);
			\node at (0.25,0.25) {\scriptsize $\Bar{\mathbf A}_4\mathbf x$};
			\node at (0.25,0.75) {\scriptsize $\Bar{\mathbf A}_3\mathbf x$};
			\node at (0.25,1.25) {$S_2$};
		\end{scope}

		\begin{scope}[scale = 3,xshift=-5]
			\draw[shaping] (0,0) rectangle (0.5,0.5);
			\draw[shaping,shade] (0,0.5) rectangle (0.5,1);
			\node at (0.25,0.25) {\scriptsize $\Bar{\mathbf A}_6\mathbf x$};
			\node at (0.25,0.75) {\scriptsize $\Bar{\mathbf A}_5\mathbf x$};
			\node at (0.25,1.25) {$S_3$};
		\end{scope}		
	\end{scope}

    \begin{scope}[xshift=450]
		\begin{scope}[scale = 3,xshift=-45]
			\draw[shaping] (0,0) rectangle (0.5,0.5);
			\draw[shaping,shade] (0,0.5) rectangle (0.5,1);
			\node at (0.25,0.25) {\scriptsize $\Bar{\mathbf A}_2\mathbf x$};
			\node at (0.25,0.75) {\scriptsize $\Bar{\mathbf A}_1\mathbf x$};
			\node at (0.25,1.25) {$S_1$};
		\end{scope}
		
		\begin{scope}[scale = 3,xshift=-25]
			\draw[shaping] (0,0) rectangle (0.5,0.5);
			\draw[shaping,shade] (0,0.5) rectangle (0.5,1);
			\node at (0.25,0.25) {\scriptsize $\Bar{\mathbf A}_4\mathbf x$};
			\node at (0.25,0.75) {\scriptsize $\Bar{\mathbf A}_3\mathbf x$};
			\node at (0.25,1.25) {$S_2$};
		\end{scope}

		\begin{scope}[scale = 3,xshift=-5]
			\draw[shaping,shade] (0,0) rectangle (0.5,0.5);
			\draw[shaping,shade] (0,0.5) rectangle (0.5,1);
			\node at (0.25,0.25) {\scriptsize $\Bar{\mathbf A}_6\mathbf x$};
			\node at (0.25,0.75) {\scriptsize $\Bar{\mathbf A}_5\mathbf x$};
			\node at (0.25,1.25) {$S_3$};
		\end{scope}		
	\end{scope}	
	\end{scope}


    \begin{scope}[yshift=-150]
    \begin{scope}
		\begin{scope}[scale = 3,xshift=-45]
			\draw[shaping,shade] (0,0) rectangle (0.5,0.5);
			\draw[shaping,shade] (0,0.5) rectangle (0.5,1);
			\node at (0.25,0.25) {\scriptsize $\Bar{\mathbf A}_2\mathbf x$};
			\node at (0.25,0.75) {\scriptsize $\Bar{\mathbf A}_1\mathbf x$};
			\node at (0.25,1.25) {$S_1$};
		\end{scope}
		
		\begin{scope}[scale = 3,xshift=-25]
			\draw[shaping,shade] (0,0) rectangle (0.5,0.5);
			\draw[shaping,shade] (0,0.5) rectangle (0.5,1);
			\node at (0.25,0.25) {\scriptsize $\Bar{\mathbf A}_4\mathbf x$};
			\node at (0.25,0.75) {\scriptsize $\Bar{\mathbf A}_3\mathbf x$};
			\node at (0.25,1.25) {$S_2$};
		\end{scope}

		\begin{scope}[scale = 3,xshift=-5]
			\draw[shaping] (0,0) rectangle (0.5,0.5);
			\draw[shaping] (0,0.5) rectangle (0.5,1);
			\node at (0.25,0.25) {\scriptsize $\Bar{\mathbf A}_6\mathbf x$};
			\node at (0.25,0.75) {\scriptsize $\Bar{\mathbf A}_5\mathbf x$};
			\node at (0.25,1.25) {$S_3$};
		\end{scope}		
	\end{scope}

    \begin{scope}[xshift=225]
		\begin{scope}[scale = 3,xshift=-45]
			\draw[shaping,shade] (0,0) rectangle (0.5,0.5);
			\draw[shaping,shade] (0,0.5) rectangle (0.5,1);
			\node at (0.25,0.25) {\scriptsize $\Bar{\mathbf A}_2\mathbf x$};
			\node at (0.25,0.75) {\scriptsize $\Bar{\mathbf A}_1\mathbf x$};
			\node at (0.25,1.25) {$S_1$};
		\end{scope}
		
		\begin{scope}[scale = 3,xshift=-25]
			\draw[shaping] (0,0) rectangle (0.5,0.5);
			\draw[shaping] (0,0.5) rectangle (0.5,1);
			\node at (0.25,0.25) {\scriptsize $\Bar{\mathbf A}_4\mathbf x$};
			\node at (0.25,0.75) {\scriptsize $\Bar{\mathbf A}_3\mathbf x$};
			\node at (0.25,1.25) {$S_2$};
		\end{scope}

		\begin{scope}[scale = 3,xshift=-5]
			\draw[shaping,shade] (0,0) rectangle (0.5,0.5);
			\draw[shaping,shade] (0,0.5) rectangle (0.5,1);
			\node at (0.25,0.25) {\scriptsize $\Bar{\mathbf A}_6\mathbf x$};
			\node at (0.25,0.75) {\scriptsize $\Bar{\mathbf A}_5\mathbf x$};
			\node at (0.25,1.25) {$S_3$};
		\end{scope}		
	\end{scope}

    \begin{scope}[xshift=450]
		\begin{scope}[scale = 3,xshift=-45]
			\draw[shaping] (0,0) rectangle (0.5,0.5);
			\draw[shaping] (0,0.5) rectangle (0.5,1);
			\node at (0.25,0.25) {\scriptsize $\Bar{\mathbf A}_2\mathbf x$};
			\node at (0.25,0.75) {\scriptsize $\Bar{\mathbf A}_1\mathbf x$};
			\node at (0.25,1.25) {$S_1$};
		\end{scope}
		
		\begin{scope}[scale = 3,xshift=-25]
			\draw[shaping,shade] (0,0) rectangle (0.5,0.5);
			\draw[shaping,shade] (0,0.5) rectangle (0.5,1);
			\node at (0.25,0.25) {\scriptsize $\Bar{\mathbf A}_4\mathbf x$};
			\node at (0.25,0.75) {\scriptsize $\Bar{\mathbf A}_3\mathbf x$};
			\node at (0.25,1.25) {$S_2$};
		\end{scope}

		\begin{scope}[scale = 3,xshift=-5]
			\draw[shaping,shade] (0,0) rectangle (0.5,0.5);
			\draw[shaping,shade] (0,0.5) rectangle (0.5,1);
			\node at (0.25,0.25) {\scriptsize $\Bar{\mathbf A}_6\mathbf x$};
			\node at (0.25,0.75) {\scriptsize $\Bar{\mathbf A}_5\mathbf x$};
			\node at (0.25,1.25) {$S_3$};
		\end{scope}		
	\end{scope}	
	\end{scope}	
    \end{scope}    
\end{tikzpicture}
	\caption{The master node can recover final solution by receiving any four subtasks completed by $S_1$, $S_2$, and $S_3$.}
	\label{fig:block}
\end{figure}

We now consider a small problem to help illustrate the
advantages of our proposed scheme. We assume there are $n=3$ workers
in the system. We choose $\mathbf G$ such that
$\mathbf
{\Bar{A}}_1=\mathbf A_1$, $\mathbf {\Bar{A}}_2= \mathbf A_2$, $\mathbf {\Bar{A}}_3 =\mathbf
A_3$, $\mathbf {\Bar{A}}_4=\mathbf A_4$, $\mathbf {\Bar{A}}_5=\mathbf A_1+\mathbf
A_2+\mathbf A_3+\mathbf A_4$, and $\mathbf {\Bar{A}}_6=\mathbf A_1+2 \mathbf
A_2+3 \mathbf A_3+4\mathbf A_4$. Each is an $m/4 \times q$
matrix. Acquiring any four of these sub-matrices is sufficient to
recover $\mathbf A$. Each worker is allocated two blocks, e.g., the
first worker gets $\mathbf {\Bar{A}}_1$ and $\mathbf {\Bar{A}}_2$. Each worker then
computes $\mathbf {\Bar{A}}_i\mathbf x$ and sends the result back to
master. In Fig.~\ref{fig:block} we illustrate all combinations of four
blocks from which the desired solution can be recovered.

In this particular example of three workers, the previous approach
of~\cite{Lee:ISIT16} can only use $k=2$ as $k<n=3$. This means that the block size in \cite{Lee:ISIT16} will be twice that of our approach. If
we compare the two approaches, the desired solutions when $k=2$ in
\cite{Lee:ISIT16} can only be obtained when two workers finish {\em
  all} their assigned work (similar to the three combinations in the
lower row of Fig.~\ref{fig:block}). In contrast, our scheme is able to
exploit the work completed by all workers and so can also recover from
the combinations of completions in the top row of
Fig.~\ref{fig:block}. As our analysis of the next section confirms,
this change results in significant acceleration.

One can infer that the higher the $k$ the larger the number of
combinations that can be used to recover the desired solution. This is
evident in the above example where we used $k=4$ (in comparison to
$k=2$). By increasing $k$ each sub-job is smaller and we can therefore
reduce the finishing time of each block. This increases the possibility of being able to exploit work performed by all processors.

\section{Matrix-Matrix Multiplication} \label{secmatrix}
In this section the objective of the master node is to compute the matrix multiplication $\mathbf A^T\mathbf B$ where $\mathbf A \in \mathbb R^{d \times q}$ and $\mathbf B \in \mathbb R^{d \times q}$. In our approach the master node decomposes $\mathbf A$ into $k$ equal sized sub-matrices $\mathbf A=[ \mathbf A_1, \mathbf A_2, \ldots, \mathbf A_k]$ where $\mathbf A_i \in \mathbb R^{d \times q/k}$. It similarly decomposes $\mathbf B$ into $\mathbf B =[\mathbf B_1, \mathbf B_2, \ldots, \mathbf B_k]$ with $\mathbf B_i \in \mathbb R^{d \times q/k}$. 
In this section, we apply the technique that we developed in the previous section to matrix-matrix multiplication using product code. We then demonstrate that tasks should be executed in a specific order by each processor to minimize the finishing time.

\subsection{Product Codes} \label{product code}
All sub-matrices $\mathbf A_i$ and $\mathbf B_j$ are further divided into $r$ smaller sub-matrices $\mathbf A_i = [\mathbf A_{i1}, \mathbf A_{i2}, \ldots, \mathbf A_{ir}]$ and  $\mathbf B_j = [\mathbf B_{j1}, \mathbf B_{j2}, \ldots, \mathbf B_{jr}]$ respectively\footnote{The reason for us to have two steps decomposition of matrices is to distinguish our approach from \cite{Lee:MATRIXISIT17}. The second step is not present at \cite{Lee:MATRIXISIT17}, therefore, each worker computes a big one tasks. However, in our scheme, each worker computes several small sub-tasks. The computation loads of each worker remain the same in both methods.}.  This creates $(kr)^2$ possible sub-computations, i.e., $\mathbf A_{ia}^T \mathbf B_{jb}$ for all $i,j \in [k]$ and $a,b \in [r]$. Note that once the master node has $\mathbf A_{ia}^T \mathbf B_{jb}$, it can recover $\mathbf A^T\mathbf B$. We arrange the $(kr)^2$  sub-computations $\mathbf A_{ia}^T \mathbf B_{jb}$ in a $kr\times kr$ array with $\mathbf A_{ia}^T \mathbf B_{jb}$ as the ($(i-1)\times r+a$,  $(j-1) \times r+b$) th element. The master node encodes each column and row using an ($n \times r$, $k \times r$) MDS code. The new coded array is similar to an $(n \times r, k \times r)^2$  product code. This creates $L=N \times R$ subtasks, where $N=n^2$ and $R=r^2$. The $L$ encoded subtasks are partitioned into $N$ arrays of size $R$, and each worker is assigned a distinct array of $R$ subtasks. Each worker sequentially works through its $R$ subtasks. At the completion of each subtask, it transmits the result to the master node.



\subsection{Order of processing}
\begin{figure}
	
	\input{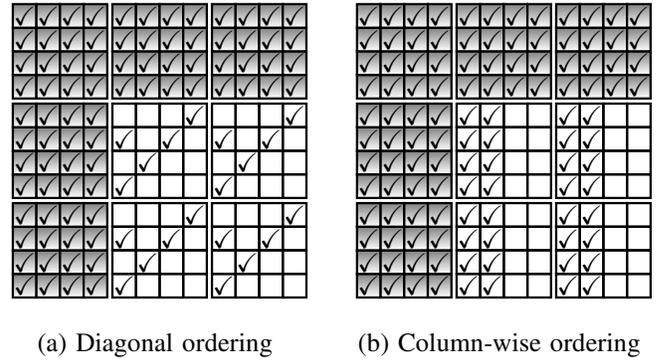}
	
	\caption{The computation time when $(a)$ all processors schedule their tasks diagonally is $1.25$ sec, while when $(b)$ all processors schedule their tasks column-wise is $2$ sec. We use a product coded scheme and have $N=9$ processors and for which $R=16$ subtasks. }
	
	\label{fig:MATMUL2}
	
\end{figure}

Through an example, we illustrate that optimizing the order in which the processors complete sub-tasks can provide a further reduction in computation time. We consider a simple distributed setup in which the master node needs to multiply two matrices  $\mathbf A=[\mathbf A_1 ,\mathbf A_2]$ and $\mathbf B =[\mathbf B_1 , \mathbf B_2]$, where $\mathbf A_i$ and $\mathbf B_j$ are $m \times q$ matrices and $k=2$, and there are $n^2 = 9$ worker nodes. We first divide each $\mathbf A_i$ and $\mathbf B_j$ into $r=4$ sub-matrices $\mathbf A_i = [\mathbf A_{i1}, \ldots ,\mathbf A_{i4}]$ and  $\mathbf B_j = [\mathbf B_{j1}, \dots, \mathbf B_{j4}]$ respectively. We then form an $8 \times 8$ array as described in~\ref{product code}. Then by encoding rows and columns of this array using $(12,8)$ MDS codes, $(n \times r)^2 = 144$ subtasks are generated. These subtasks are arranged in a $12 \times 12$ array, as shown in Fig.~\ref{fig:MATMUL2}.
We assign $r^2=16$ subtasks to each processor, as depicted by the $4 \times 4$ blocks in Fig.~\ref{fig:MATMUL2}. For illustrative purpose we assume that the four white processors are stragglers and complete their tasks in 4 sec. We assume that the other five gray processors are non-stragglers and finish their tasks after 1 sec. In one approach all processors perform their tasks according to a diagonal schedule illustrated in Fig.~\ref{fig:MATMUL2}a. In the second, they follow the column-wise scheduling depicted in Fig.~\ref{fig:MATMUL2}b. If the diagonal schedule is used, the master node can complete the matrix-matrix multiplication $\mathbf A^T\mathbf B$ in 1.25 sec. If the column-wise schedule is used, 2 sec are needed. In both approaches the completed subtasks at the end of computation of $\mathbf A^T\mathbf B$  are marked as checks boxes. Later in the numerical section, we further evaluate the order of processing in detail.

\section{Theoretical analysis} \label{theoritical}

In this section, we evaluate the performance of our proposed scheme for vector-matrix multiplication based on MDS codes. We propose a novel approach based on the amount of work completed in some fixed time to study performance. 

\begin{figure}
	\centering
	\includegraphics[keepaspectratio,width=1\columnwidth]{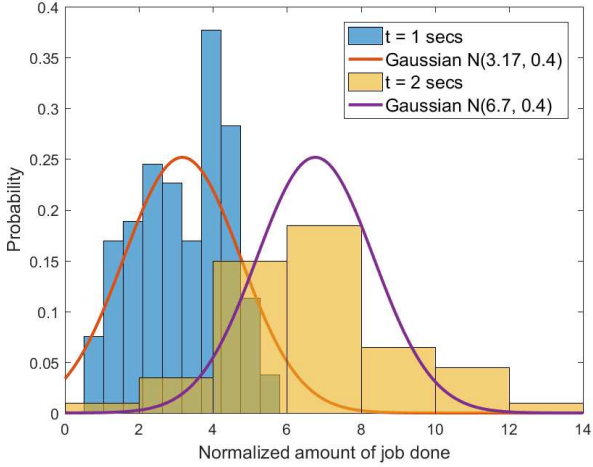}
	\caption{The distribution of the number of jobs completed by a
          given time.}
	\label{fig:distribution}
\end{figure}

Our approach is to quantify processing time based on
the amount of work completed by each worker. First we find a suitable distribution that captures the amount of work completed by each worker in a given amount of time. Let $V_i^t$ be a random variable that denotes the amount of work completed
by the $i$-th worker by time $t$ and let $p_{i}(v_i^t)$ be the
probability distribution of $V_i^t$. For this analysis we consider
$p_i(v_i^t)$ to be distributed as a discrete Gaussian random
variable. 
In Fig.~\ref{fig:distribution} we plot the processing workload of a
randomly busy processor, selected randomly from a number of
  processors running on EC2. In this experiment we specified fixed
times ($t=1$ and $t=2$ seconds), and allowed the processor to compute a
number vector-matrix multiplications until the specified time. We then
counted the number of jobs (vector-matrix multiplications)
completed. We observe the distribution of number of jobs completed is roughly
Gaussian. We note that we consider the Gaussian distribution for only positive values because the amount of work completed by each worker by time $t$ cannot be negative. One can notice that the average ($\gamma_i^t$) number of
jobs completed is a function of $t$ and that it approximately doubles
when $t$ is doubled. The variation (${\sigma_i^t}^2$) also increases slightly with $t$. Based on these
assumption, we assume the distribution of $v_i^t$ to be
\begin{align}
p_i(v_i^t|v_i^t\in [l_i]) = \frac{1}{c_i\sqrt{2\pi{\sigma_i^t}^2}} e^{-\frac{(v_i^t-\gamma_i^t)^2}{2{\sigma_i^t}^2}}
\end{align} 
where $[l_i]=\{0,\ldots l_i\}$ and $c_i = \sum_{j=0}^{l_i}  \frac{1}{\sqrt{2\pi{\sigma_i^t}^2}} e^{-\frac{(j-\gamma_i^t)^2}{2{\sigma_i^t}^2}} $.
We need to determine the probability that the master receives $k$
distinct blocks by time $t$ (so that the overall job has completed by
time $t$). Let us define the random variable $Z_t$ to be
\begin{align}
\label{eqn:sum.Gaussian}
Z_t = \sum_{i=1}^{n} V_i^t.
\end{align}
As the workers are assumed to be independent, the number of jobs
completed by the workers are also jointly independent and therefore
$Z_t$ is a sum of independent discrete Gaussian distribution, which is
equal to a discrete Gaussian with distribution
\begin{align}
p\left(z_t|z_t\in \left[L\right]\right) = \frac{1}{\sqrt{2\pi{\sigma_z^t}^2}} e^{-\frac{(z_t-\gamma_z^t)^2}{2{\sigma_z^t}^2}}
\end{align}
where 
\begin{align*}
\gamma_z^t = \sum_{i=1}^{n} \frac{\gamma_i^t}{c_i}, \ \ 
{\rm and} \ \ 
\sigma_z^t=\sqrt{\left(\sum_{i=1}^{n} {\frac{{\sigma_i^t}^2}{{c_i}^2}}\right)}.
\end{align*}
We can now find the probability that the master node is able to
collect $k$ distinct blocks by time $t$:
\begin{align}
\label{eqn:susscess.probability}
\text{Pr}[Z_t\geq k]   = \sum_{z_t=k}^{L} p_i(z_t|z_t\in \mathbb Z). 
\end{align}
In contrast to our approach, the
lack of sub-blocking in the earlier literature~\cite{Lee:ISIT16,
  Avestimehr:ISIT17} restricted the $V_i^t$ to be in the set
$\{0,l_i\}$. The option for the $V_i^t$ to take on a larger set of
values gives our scheme a material advantage.

\begin{figure}
	\centering
	\includegraphics[keepaspectratio,width=1\columnwidth]{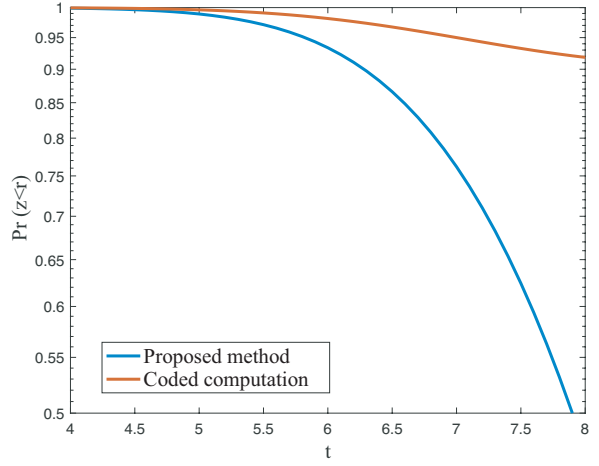}
	\caption{The probably of {\em non}-completion by a given time versus $t$, $\text{Pr}[Z_t<k]$.}
	\label{fig:zlessr}
\end{figure}

Fig.~\ref{fig:zlessr} plots the probability that
the master node does not acquire at least $k$ blocks as a function of time, i.e,
$\text{Pr}[Z_t< k] = 1-\text{Pr}[Z_t\geq k]$. This figure
plots~\eqref{eqn:susscess.probability} when $m=1000, n=10, L=100,
k=40, \gamma_i^t=0.5t$ and $\sigma_i^t=2$.  It can be
observed that, in our proposed scheme, the probability of not
finishing the computation task by time $t$ decays much more
quickly than for the scheme of~\cite{Lee:ISIT16}.

\section{Numerical Evaluation} \label{simulation}
In this section several Monte Carlo simulations are  presented to estimate the finishing time of different schemes. We compare our proposed scheme that exploits stragglers to the frequently-used approach of completely ignoring stragglers. In each trial of our simulations we generate $N$ independent exponential random numbers with mean $ \frac{1}{\lambda} =1$. These $N$ numbers are denoted by $ T_1, T_2, \ldots, T_N $, i.e., $T_i$ is the time to complete all subtasks. We consider  $\frac{j T_i}{r^2}$ as the finishing time of the $j-$th subtask of the $i-$th processor where $j \in \{ 1,2,3, \ldots ,r^2 \} $ and $i \in \{ 1,2,3, \ldots , N\}$.  In all simulations we set $k=20$.

\begin{figure}
	
	\tikzset{every mark/.append style={scale=1}}
		\begin{tikzpicture}[scale=0.8]
	\begin{axis}[
	height=8cm,
	width=10cm,
	grid=major,
	xlabel={\# of Processors, $N$},
	ylabel={Expected Finishing Time, $E(T)$},
	legend style={nodes=right},
	axis on top,xmin=600, xmax=2400, ymin=0, ymax=1.6]
	\addlegendentry{MDS code (multiple MDS), $r=1$ \cite{Lee:ISIT16, Lee:MATRIXISIT17}}
	\addplot [line width=0.5mm, color=blue, solid, every mark/.append style={solid, fill=blue},mark=otimes*] coordinates {
		(600,	1.586689)
		(800,	1.009521)
		(1000,	0.727410)
		(1200,	0.590140)
		(1400,	0.487718)
		(1600,	0.419143)
		(1800,	0.376451)
		(2000,	0.323725)
		(2200,	0.287811)
		(2400,	0.259282)

	};

	\addlegendentry{Product code, $r=1$ \cite{Lee:MATRIXISIT17}}
	\addplot [line width=0.5mm, color=black, solid, every mark/.append style={solid, fill=black},mark=asterisk] coordinates {
		(600,	1.329081)
		(800,	0.867461)
		(1000,	0.694460)
		(1200,	0.597659)
		(1400,	0.502904)
		(1600,	0.438000)
		(1800,	0.405806)
		(2000,	0.384491)
		(2200,	0.345983)
		(2400,	0.329606)
	};

	\addlegendentry{Lower bound (single MDS), $r=1$ \cite{Lee:ISIT16}}
	\addplot [line width=0.5mm, color=red, solid, every mark/.append style={solid, fill=red},mark=square] coordinates {
		(600,	1.095849)
		(800,	0.693977)
		(1000,	0.506180)
		(1200,	0.410906)
		(1400,	0.335401)
		(1600,	0.288049)
		(1800,	0.252853)
		(2000,	0.224009)
		(2200,	0.196420)
		(2400,	0.180747)
	};

	\addlegendentry{MDS code (multiple MDS), $r=2$}
	\addplot [line width=0.5mm, color=blue, dashed, every mark/.append style={solid, fill=blue},mark=otimes*] coordinates {
		(600,	0.948257)
		(800,	0.565460)
		(1000,	0.386358)
		(1200,	0.308176)
		(1400,	0.253903)
		(1600,	0.217247)
		(1800,	0.187097)
		(2000,	0.160693)
		(2200,	0.143488)
		(2400,	0.130178)
	};

	\addlegendentry{Product code, $r=2$           }
	\addplot [line width=0.5mm, color=black, dashed, every mark/.append style={solid, fill=black},mark=asterisk] coordinates {
		(600,	0.804400)
		(800,	0.498468)
		(1000,	0.390341)
		(1200,	0.323052)
		(1400,	0.275638)
		(1600,	0.238535)
		(1800,	0.217974)
		(2000,	0.205333)
		(2200,	0.185406)
		(2400,	0.175255)
	};

	\addlegendentry{Lower bound (single MDS), $r=2$}
	\addplot [line width=0.5mm, color=red, dashed, every mark/.append style={solid, fill=red},mark=square] coordinates {
		(600,	0.629497)
		(800,	0.372785)
		(1000,	0.263992)
		(1200,	0.210107)
		(1400,	0.170733)
		(1600,	0.144137)
		(1800,	0.126084)
		(2000,	0.111597)
		(2200,	0.097571)
		(2400,	0.089472)
	};		
	\end{axis}
	\end{tikzpicture} 
	
	\caption{The expected finishing time $E(T)$ vs. number of processors ($N$) for different number of subtasks ($r^2$).}
	\label{fig:1}
\end{figure}
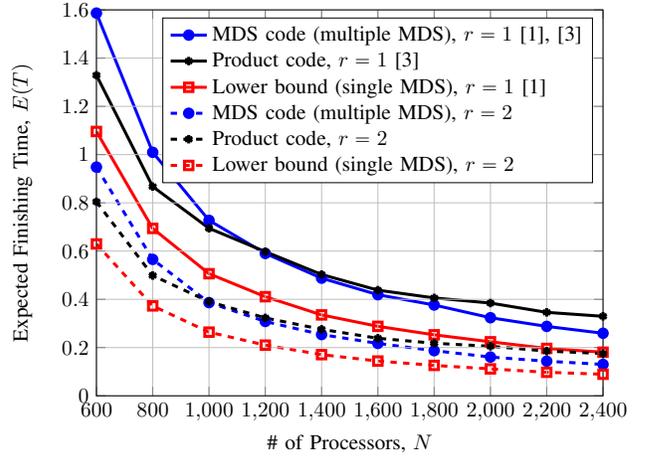

In Fig.~\ref{fig:1} the expected finishing times of the multiple ($\frac{N}{k} \times r$, $k \times r$) MDS-coded, ($\lfloor \sqrt{N} \rfloor \times r$, $k \times r$)$^2$ product-coded, and single ($ N \times r^2$, $(k \times r)^2$) MDS-coded schemes are plotted for $N \in \{ 600,800,1000,...,2400\}$. We plotted results for both $r=1$ (equivalent to \cite{Lee:ISIT16,Lee:MATRIXISIT17}) and $r=2$. Fig.~\ref{fig:1} shows that our method significantly reduces the finishing time. The gain is a result of increasing $r=1$ to $r=2$ such that each worker computes $r^2=4$ subtasks sequentially rather than $r^2=1$ (big) subtask as in \cite{Lee:MATRIXISIT17}. 

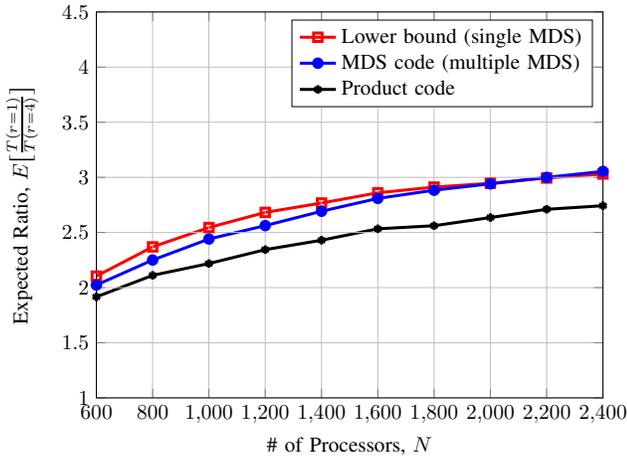
\begin{figure}
	\tikzset{every mark/.append style={scale=1}}
		\begin{tikzpicture}[scale=0.8]
	\begin{axis}[
	height=8cm,
	width=10cm,
	grid=major,
	xlabel={\# of Processors, $N$},
	ylabel={Expected Ratio, $E\big [ \frac{T(r=1)}{ T(r=4)}\big ]$},
	legend style={nodes=right},
	axis on top,xmin=600, xmax=2400, ymin=1, ymax=4.5]
	
	\addlegendentry{Lower bound (single MDS)}
	\addplot [line width=0.5mm, color=red, solid, every mark/.append style={solid, fill=red},mark=square] coordinates {
		
		(600,	2.104487)
		(800,	2.369613)
		(1000,	2.545070)
		(1200,	2.682093)
		(1400,	2.767178)
		(1600,	2.860118)
		(1800,	2.911650)
		(2000,	2.946091)
		(2200,	2.994301)
		(2400,	3.030343)

	};

	\addlegendentry{MDS code (multiple MDS)}
	\addplot [line width=0.5mm, color=blue, solid, every mark/.append style={solid, fill=blue},mark=otimes*] coordinates {
		(600,	2.023743)
		(800,	2.250417)
		(1000,	2.440956)
		(1200,	2.562294)
		(1400,	2.692575)
		(1600,	2.809093)
		(1800,	2.883322)
		(2000,	2.941230)
		(2200,	2.999739)
		(2400,	3.054129)
	};

	\addlegendentry{Product code           }
	\addplot [line width=0.5mm, color=black, solid, every mark/.append style={solid, fill=black},mark=asterisk] coordinates {
		(600,	1.916085)
		(800,	2.111464)
		(1000,	2.218493)
		(1200,	2.344548)
		(1400,	2.429854)
		(1600,	2.533437)
		(1800,	2.561288)
		(2000,	2.635739)
		(2200,	2.710257)
		(2400,	2.742838)
	};
	
	\end{axis}
	\end{tikzpicture}
	\caption{The ratio of expected times of our scheme at $r=4$ to \cite{Lee:MATRIXISIT17} vs. the number of processors ($N$).}
	\label{fig:2}
\end{figure}

The ratio of improvement between $r=1$ and $r=4$ is depicted in Fig.~\ref{fig:2} for $N \in \{600,800,1000,...,2400\}$. Fig.~\ref{fig:2} shows that by dividing the main task into $r^2 = 16$ subtasks is at least twice as good as compared to $r=1$ \cite{Lee:MATRIXISIT17} and reaches three times for larger number of workers.

The impact of increasing $r$ on the average finishing time is shown in Fig.~\ref{fig:3}. The improvement from $r=1$ to $r=2$ is significant. For $r > 4$ the further improvement, while positive, is not very significant. Therefore, excessively increasing in the number of subtasks is not logical due to the additive complexity incurred.

\begin{figure}
	\tikzset{every mark/.append style={scale=1}}
		\begin{tikzpicture}[scale=0.8]
	\begin{axis}[
	height=8cm,
	width=10cm,
	grid=major,
	xlabel={$r$, \# of subtasks = $r^2$},
	ylabel={Expected Finishing Time, $E(T)$},
	legend style={nodes=right},
	axis on top,xmin=1, xmax=8, ymin=0.6, ymax=1.4]
	\addlegendentry{Product code with column ordering}
	\addplot [line width=0.5mm, color=black, dotted, every mark/.append style={solid, fill=black},mark=asterisk] coordinates {
		(1,	1.330775)
		(2,	0.851905)
		(3,	0.791339)
		(4,	0.761816)
		(5,	0.760941)
		(6,	0.748919)
		(7,	0.743432)
		(8,	0.746414)
	};

	\addlegendentry{Product code with random ordering}
	\addplot [line width=0.5mm, color=black, solid, every mark/.append style={solid, fill=black},mark=asterisk] coordinates {
		(1,	1.330775)
		(2,	0.808114)
		(3,	0.717115)
		(4,	0.687596)
		(5,	0.685714)
		(6,	0.674491)
		(7,	0.666060)
		(8,	0.668216)
	};

	\addlegendentry{Product code with diagonal ordering}
	\addplot [line width=0.5mm, color=black, dashed, every mark/.append style={solid, fill=black},mark=asterisk] coordinates {
		(1,	1.330775)
		(2,	0.781518)
		(3,	0.708904)
		(4,	0.687435)
		(5,	0.689134)
		(6,	0.681433)
		(7,	0.677304)
		(8,	0.680027)
	};
	
	\end{axis}
	\end{tikzpicture}
	\caption{The expected finishing time $E(T)$ vs. $r$ for different processing orders. We fixed $N=600$. }
	\label{fig:4}
\end{figure}
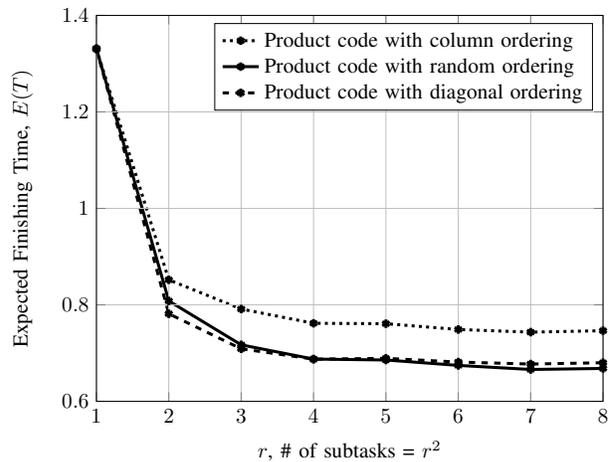

Fig.~\ref{fig:4} illustrates the average finishing time when the order of processing is changed. In this figure we set $N=600$ and vary $r \in \{1,2,3,...,8\}$. It is observed in Fig.~\ref{fig:4} that diagonal order of processing closely matches the random ordering. Further, column-wise (or row-wise) processing order is a bad choice.

\begin{figure}
	\tikzset{every mark/.append style={scale=1}}
		\begin{tikzpicture}[scale=0.8]
	\begin{axis}[
	height=8cm,
	width=10cm,
	grid=major,
	xlabel={$r$, \# of Subtasks = $r^2$},
	ylabel={Expected Finishing Time, $E(T)$},
	legend style={nodes=right},
	axis on top,xmin=1, xmax=8, ymin=0, ymax=2]
	\addlegendentry{MDS code (multiple MDS)}
	\addplot [line width=0.5mm, color=blue, solid, every mark/.append style={solid, fill=blue},mark=otimes*] coordinates {
		(1,	1.597205)
		(2,	0.932222)
		(3,	0.824177)
		(4,	0.787973)
		(5,	0.781439)
		(6,	0.749668)
		(7,	0.738722)
		(8,	0.734145)
	};

	\addlegendentry{Product code           }
	\addplot [line width=0.5mm, color=black, solid, every mark/.append style={solid, fill=black},mark=asterisk] coordinates {
		(1,	1.334650)
		(2,	0.779835)
		(3,	0.713141)
		(4,	0.698728)
		(5,	0.691825)
		(6,	0.688828)
		(7,	0.680401)
		(8,	0.681183)
	};

	\addlegendentry{Lower bound (single MDS)}
	\addplot [line width=0.5mm, color=red, solid, every mark/.append style={solid, fill=red},mark=square] coordinates {
		(1,	1.098500)
		(2,	0.629738)
		(3,	0.548799)
		(4,	0.525005)
		(5,	0.512028)
		(6,	0.509786)
		(7,	0.499635)
		(8,	0.495645)
	};
	
	\end{axis}
	\end{tikzpicture}

	\caption{The expected finishing time  $E(T)$ vs. $r$ for $N=600$ processors. }
	\label{fig:3}
	
\end{figure}
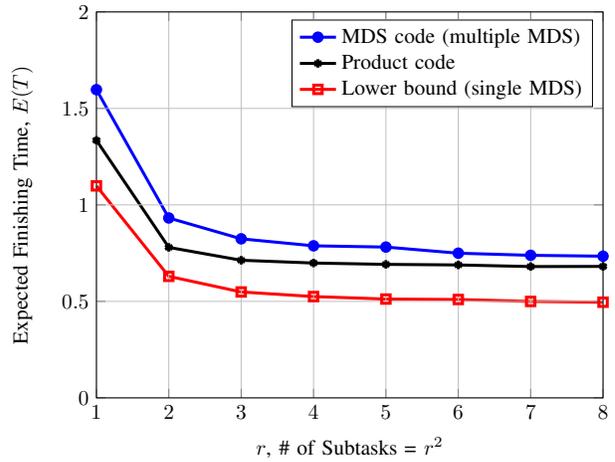

\section{Conclusion}
In this paper we have proposed a method to exploit the work completed by stragglers in distributed coded computation. We first applied our method to vector-matrix multiplication based on MDS codes. The main idea is to assign a large number of small MDS-coded jobs to workers, rather than to assign each worker a single (larger) job. By allowing workers to work on small jobs, workers can transmit back each partial solution as they complete each small job. Through these changes, we realize significant acceleration in comparison to previous approaches.  We then extend our work to matrix-matrix multiplication. By selecting a suitable order of processing we achieved additional improvement in finishing time. We analyzed our scheme for MDS coded vector-matrix multiplication. The simulations show more than a factor of two improvement in the expected finishing time. 
\bibliographystyle{IEEEtran} 
\bibliography{reference}
\end{document}